\documentclass[12pt]{iopart}
\usepackage{graphicx}
\usepackage{graphics}
\usepackage{latexsym}
\usepackage{amsmath, amsthm, amssymb}
\usepackage{xcolor}
\usepackage{hyperref}

\begin{document}

\title[]{Steady state and relaxation dynamics of run and tumble particles in contact with a heat bath}

\author{R. K. Singh}
\address{Department of Biomedical Engineering, Ben-Gurion University
of the Negev, Be'er Sheva 85105, Israel}
\ead{rksinghmp@gmail.com}

\author{Oded Farago}
\address{Department of Biomedical Engineering, Ben-Gurion University
of the Negev, Be'er Sheva 85105, Israel}
\ead{ofarago@bgu.ac.il}

\begin{abstract}
We study the relaxation dynamics of a run and tumble particle in a one-dimensional piecewise linear potential 
$U(x)=b|x|$, from delta-function initial conditions at $x=0$ to steady state. In addition to experiencing active telegraphic noise, 
the particle is in contact with a heat bath at temperature 
$T$ that applies white thermal noise.
We find that the position distribution of the RTP is described by a sum of two distributions ("modes"), each of which of the form
$P(x,t\to\infty)\sim e^{-\lambda_i|x|}$ ($i=1,2$) at steady state. The two modes are dynamically coupled: At very short times ($t\to 0$), each mode stores half of the
probability, and exhibits thermal diffusive spreading with a Gaussian profile. With progressing time and evolution toward steady state, the partition of probability  between the modes becomes increasingly uneven and, depending on the model parameters, the mode with the smaller value of $\lambda_i$ may carry 
an overwhelming majority of the probability. Moreover, we identify that the characteristic relaxation time of each mode is $\tau_i=(\lambda_i^2T)^{-1}$, which implies that the minority mode also relaxes much faster than the dominant one. A more detailed analysis reveals that $\tau_i$ is characteristic of the mode relaxation only close to the origin at the core of the distribution, while further away it increases linearly with $|x|$ as if a relaxation front is propagating at constant speed $v_i^*=2\sqrt{T/\tau_i}$ in the system. 
The rate of non-equilibrium entropy production can be related to the two-mode splitting of the probability distribution and be expressed in terms of their correlation-lengths $\lambda_i$ and their contributions to the steady state distribution.  
\end{abstract}

\newcommand{\fr}{\frac}
\newcommand{\tl}{\tilde}
\newcommand{\dd}{\partial}
\newcommand{\lr}{\langle}
\newcommand{\rl}{\rangle}

\section{Introduction}
Active particles use energy from the environment to fuel their motion which
is inherently out of equilibrium
\cite{romanczuk2012active,marchetti2013hydrodynamics,ramaswamy2017active,
bechinger2016active,fodor2018statistical}.
Activity dominates the dynamics of, e.g., molecular motors
\cite{backouche2006active,mizuno2007nonequilibrium,toyota2011non,
stuhrmann2012nonequilibrium}, transport within cells \cite{wilhelm2008out,
ahmed2015active,cates2012diffusive}, bird flocking
\cite{vicsek1995novel,toner2005hydrodynamics}, to mention a few.
The ubiquity of active particles has prompted investigations of their
properties also from the perspective of applications, like designing phoretic
swimmers \cite{golestanian2007designing}, acoustic and harmonic trapping
of active systems \cite{takatori2016acoustic,dauchot2019dynamics}, and
engineering self-propelled robots \cite{deblais2018boundaries}.

Bacterial dynamics in fluids constitutes yet another example of non-equilibrium 
active motion. Movement of bacteria in fluids is characterized by phases of runs in which
the bacteria travels in straight lines, separated by tumbling events marking
the directional changes \cite{di2010bacterial,sokolov2010swimming,
saragosti2011directional,sidortsov2017role,taktikos2013motility,goral2022frustrated}.
These dynamical aspects of bacterial motion have been explored 
within the model of run and tumble particles (RTPs), which is one of the 
more extensively studied models in non-equilibrium statistical physics 
\cite{solon2015active,cates2013active,elgeti2015run,ezhilan2015distribution,
angelani2017confined,villa2020run,mori2020universal,mori2020prl,Frydel_2021,angelani2023one}.
In the simplest setting, an RTP moves with a fixed speed, undergoing random tumbling
events at a constant rate. The resulting motion
is described by the telegrapher's equation \cite{masoliver1996finite,
porra1997telegrapher,weiss2002some}, which unlike the diffusion equation for
Brownian motion, is second order in both space and time. This leads to many interesting 
physical properties, notable being the motion confined within a spatiotemporal 
light cone \cite{klafter2011first}. Nevertheless, at times much larger than the 
mean tumbling time and in free space, an RTP exhibits diffusive behavior similar to a 
Brownian particle. 

The differences in the dynamical properties of the RTP and a Brownian particle become
apparent when their respective motions are subject to confinement by an external 
potential. Generally speaking, the steady state of an RTP in a confining potential
is non-Boltzmannian \cite{dhar2019run,sevilla2019stationary,
le2020velocity}. Moreover, if the associated restoring force exceeds the active force as, 
for instance, in the case of a harmonic potential, the steady state of the RTP is 
defined only over a finite support \cite{dhar2019run,smith2022nonequilibrium}. 
This is in sharp contrast to a Brownian motion in the same
potential, for which the steady state has an infinite support
\cite{risken1996fokker}. The intrinsic
activity of RTPs also renders their first passage properties complex 
and counter to the intuition based on the first passage
properties of Brownian particles \cite{gueneau2023optimal,gueneau2024run}.

In most cases of practical interest, the motion of RTPs often takes place 
at a finite temperature \cite{gachelin2014collective,wioland2016directed}. 
In the literature, however, most studies focus on RTP dynamics under the influence of 
a telegraphic (dichotomous) noise only, and the influence of an additional white thermal 
noise  has been relatively less explored 
\cite{tucci2022modeling,wexler2020dynamics,garcia2021run,
frydel2022positing,arcobi2023continuous,le2021stationary,smith2022exact, nandi2024optimizing}. 
The significant differences between the statistical behavior 
of RTPs and Brownian particles makes it imperative to study the dynamical properties
of RTPs that are also subject to both types of noises.  
The prime question is whether the coupling to a heat bath modifies the dynamical
properties in a nontrivial way, or just increases the fluctuations in position
due to the additional thermal noise. This question is particularly relevant for understanding physical systems such as bacteria, whose motion can be effectively described by the run-and-tumble particle (RTP) model, but which, in reality, operate at finite temperatures. A predictable consequence of introducing white thermal noise into the model is that the particle's motion will no longer be restricted to a finite spatial domain. In this paper, we present a rare example where (i) the steady-state distribution (SSD) can be computed exactly, and (ii) the dynamical evolution toward the steady state can be analyzed in Laplace space and approximated in the time domain. Our findings highlight that the interplay between active and thermal noise can lead to intricate dynamical behavior.

The dynamics of an RTP in a one-dimensional potential $U(x)$ and in contact with a heat
bath is described by the Langevin equation:
\begin{align}
\label{lang}
\frac{dx}{dt} = f(x(t)) +\sigma(t) + \eta(t),
\end{align}
where $f(x) = -dU(x)/dx$ is the force acting on the particle, while $\sigma(t)$ and $\eta(t)$ are 
active telegraphic (dichotomous) and thermal noises, respectively. Note that we set the mobility $\mu$  to unity in Eq.~(\ref{lang}), 
so that the forces on the r.h.s.~of Eq.~(\ref{lang}) 
are expressed in units  of velocity (and the conversion back to units of force is done via: ${\rm force}=\mu^{-1}\cdot{\rm velocity}$). The active noise switches between the two values $\sigma(t)=\pm v_0$ 
at a fixed rate $r$,  i.e., the time between consecutive switches is drawn from an exponential distribution 
with mean running time $r^{-1}$. The thermal noise is a white Gaussian noise with zero mean and 
correlation $\langle \eta(t) \eta(s) \rangle= 2T\delta(t-s)$, where $T$ denotes the temperature of 
the heat bath.  Here, we adopt the common convention of setting Boltzmann constant $k_B$ to unity. In this notation, diffusion coefficient $D$ is expressed in units of the corresponding temperature, as defined by the Einstein relation: $D=\mu k_BT$. Interestingly, Langevin equation (\ref{lang}) describing the motion of an RTP in 
contact with a heat bath, also describes the motion of a Brownian particle in a dichotomously fluctuating 
potential $U(x)\pm v_0(t)x$. The latter problem was considered in a 1993 paper by Bier and Astumian 
(BA)~\cite{bier1993matching} who studied the mean escape time from a fluctuating potential well. 
Here, we exploit the equivalence between the two problems to derive expressions for the SSD 
of an RTP moving in an external potential and subject to thermal noise. We extend BA approach to study also the 
corresponding time-dependent Fokker-Planck (FP) equation, analyze the relaxation of an initial delta-function distribution   
to the SSD, and unravel the interplay between the active and thermal noises. As in ref.~\cite{bier1993matching}, we consider a piecewise linear potential of the form $U(x)=b|x|$,  which is also relevant to the study of active particle sedimentation, where the steady-state distribution (SSD) is determined by the solution of the Mathieu equation~\cite{mathieu}. Within the context of the RTP model, the linear potential is appealing for at least two reasons. First, it allows for an exact analytical solution, namely a derivation of a closed-form expression of the SSD, because the total force acting on the particle takes only two values: $(v_0-b)$ when moving outward of the origin, and $(v_0+b)$ when moving toward the origin. Second, unlike a harmonic potential (and more generally, any potential of the form $U(x)\sim|x|^a$ with $a>1$), the piecewise linear potential has infinite support even in the absence of thermal noise, as long as $v_0>b$. In fact, the SSD of such an RTP has a form which is similar to the equilibrium Boltzmann distribution \cite{dhar2019run,farago2024confined}: 
\begin{align}
    P(x)\sim e^{-b|x|/T^*},
    \label{expssd}
\end{align}
with $T^*=(v_0^2-b^2)/2r$. However, as will be shown below, the SSD of a particle following Langevin equation (\ref{lang}) does {\em not}\/ have the form of Eq.~(\ref{expssd}) with temperature $T+T^*$, but rather follows a double exponential form. A single exponential form is obtained only in the limit when both  $v_0 \to \infty$ and $r \to \infty$ such that the ratio $T_{\rm ac}=v^2_0/2r$ is fixed. In this limit, the active noise approaches a white Gaussian noise with mean zero and correlation
$\langle \sigma(t) \sigma(s) \rangle = 2T_{ac}\delta(t-s)$, where $T_{\rm ac}$ is the active "temperature" (or the associated diffusion coefficient of the active noise - see comment above about the correspondence between temperature and diffusion coefficient). Indeed, for a flat potential ($b=0$), $T_{\rm ac}$ is the large-time diffusion coefficient of an RTP not coupled to a heat bath, for any value of $v_0$ and $r$.   

The paper is organized as follows: In Section~\ref{sec:ssd} we study the long
time behavior of the distribution, $P(x,t)$, and derive the expression for the SSD,
$P(x)$. In the following Section~\ref{secPxt} we obtain the expression for the temporal
evolution of the position distribution in Laplace space. An exact inversion of the function from Laplace space to the time domain is not possible. A closed-form analytical approximation is derived in Section~\ref{secNp}, where we also present Langevin Dynamics simulation results to analyze the relaxation dynamics to steady state. In section~\ref{entropy} we briefly discuss and rate of entropy production at steady state. We summarize and discuss our main findings in section~\ref{secSS}

\section{Steady state of the RTP}
\label{sec:ssd}
Let $p_{\pm}(x,t)$ denote the distributions of the
RTP to be at location $x$ at time $t$ with $\sigma(t)=\pm v_0$. Then the evolution of
the distributions can be described by the set of coupled FP equations \cite{risken1996fokker,
gardiner1985handbook}:
\begin{subequations}
\label{fpe}
\begin{align}
\dd_t p_+(x,t) &= -\dd_x [(f(x)+v_0)p_+(x,t)]
- r p_+(x,t) + r p_-(x,t) + T \dd_{xx} p_+(x,t),\\
\dd_t p_-(x,t) &= -\dd_x [(f(x)-v_0)p_-(x,t)]
+ r p_+(x,t) - r p_-(x,t) + T \dd_{xx} p_-(x,t).
\end{align}
\end{subequations}
The respective SSDs, $p_{\pm}(x)$, are obtained by solving the set of ordinary differential
equations:
\begin{subequations}
\label{fpe_ss}
\begin{align}
0 = &-\fr{d}{dx}[(f(x)+v_0)p_+(x)] - rp_+(x) + rp_-(x)
+ T\fr{d^2}{dx^2}p_+(x),\\
0 = &-\fr{d}{dx}[(f(x)-v_0)p_-(x)] + rp_+(x) - rp_-(x)
+ T\fr{d^2}{dx^2}p_-(x).
\end{align}
\end{subequations}
Symmetry of the problem implies that the above set of differential equations only needs
to be solved in the region $x \geq 0$ and the solution for the region $x \leq 0$
follows by symmetry, $p_{\pm}(-x)=p_{\mp}(x)$. Furthermore, the SSD $P(x)=p_+(x)+p_-(x)$.

Let us now focus on the region $x \geq 0$, which implies that $f(x) = -b$. Following~\cite{bier1993matching}, we
define
\begin{subequations}
\label{s0d0}
\begin{align}
P_0(x) &= p_+(x) + p_-(x),\\
P_1(x) &= \fr{dP_0}{dx},\\
D_0(x) &= p_+(x) - p_-(x),\\
D_1(x) &= \fr{dD_0}{dx},
\end{align}
\end{subequations}
where $P_0(x)=P(x)$ is the SSD of the RTP. Using these definitions, 
we can write the set of Eqs.~(\ref{fpe_ss}), in the following form
\begin{align}
\label{mat3_ss}
&\fr{d}{dx}P_0 = P_1~\text\nonumber \\
&\fr{d}{dx}
\begin{pmatrix}
P_1\\
D_0\\
D_1\\
\end{pmatrix}
= \underbrace{\begin{pmatrix}
-\fr{b}{T} & 0 & \fr{v_0}{T}\\
0 & 0 & 1\\
\fr{v_0}{T} & \fr{2r}{T} & -\fr{b}{T}
\end{pmatrix}}_{M_3}
\begin{pmatrix}
P_1\\
D_0\\
D_1\\
\end{pmatrix}.
\end{align}
The structure of the above equations implies that the solution $(P_1~D_0~D_1)^\top$,
where $\top$ denotes transpose, is determined by the eigenvalues of the matrix $M_3$,
with the eigenvalues $\lambda_k \in \mathbb{R}~\forall~k = 1,2,3$ as absence of any
boundaries forbids oscillatory solutions. Furthermore
\begin{subequations}
\begin{align}
\text{tr}~M_3 &= -\fr{2b}{T} < 0,\\
\text{det}~M_3 &= \fr{2rb}{T^2} > 0,
\end{align}
\end{subequations}
with "tr" and "det" representing, respectively, the trace and determinant of the matrix $M_3$. This implies that
two of the eigenvalues are negative and one is positive. The eigenvalues are determined
by the solution of the equation $\text{det}(M_3 - \lambda \mathbb{I}_3) = 0$, where
$\mathbb{I}_3$ is the $3\times 3$ identity matrix. The solution of the resulting cubic equation
\begin{align}
\label{eig3}
\lambda^3 + \fr{2b}{T}\lambda^2 - \Big(\fr{v^2_0}{T^2} + \fr{2r}{T}
- \fr{b^2}{T^2}\Big)\lambda - \fr{2rb}{T^2} = 0,
\end{align}
is well known \cite{nickalls2006viete,zucker200892}:
\begin{align}
\label{lmb_k}
\lambda_k = -\fr{2b}{3T} + 2\sqrt{-\fr{q_1}{3}}\cos\Big[\fr{1}{3}\cos^{-1}\Big(\fr{3q_2}
{2q_1}\sqrt{-\fr{3}{q_1}}\Big)-\fr{2\pi k}{3}\Big],
\end{align}
for $k = 0, 1, 2$, with $q_1 = -\fr{v^2_0}{T^2}-\fr{b^2}{3T^2}-\fr{2r}{T}$ and
$q_2 = \fr{2bv^2_0}{3T^3}-\fr{2br}{3T^2}-\fr{2b^3}{27T^3}$. Among the three eigenvalues,
we only choose the ones which are negative, as inclusion of the positive eigenvalue would lead
to diverging solutions at large $x$ (recall that we solve for $x\geq 0$). 

Let $\lambda_1$ and $\lambda_2$ be the {\em absolute values}\/ of two negative eigenvalues of the matrix $M_3$, then
\begin{subequations}
\label{s1d0d1}
\begin{align}
P_1(x) &= -A_1 e^{-\lambda_1 x} - A_2e^{-\lambda_2 x},\\
D_0(x) &= \fr{b-\lambda_1 T }{v_0 \lambda_1} A_1 e^{-\lambda_1 x}
+ \fr{b-\lambda_2 T}{v_0 \lambda_2} A_2 e^{-\lambda_2 x},\\
D_1(x) &= -\fr{b-\lambda_1 T }{v_0} A_1 e^{-\lambda_1 x}
- \fr{b-\lambda_2 T}{v_0} A_2 e^{-\lambda_2 x}.
\end{align}
\end{subequations}
Because $\fr{d}{dx}P_0 = P_1$, and because of the symmetry $P_0(x)=P_0(-x)$, we find that the SSD 
is given by
\begin{align}
\label{px_ss}
P(x)=P_0(x) = \fr{A_1}{\lambda_1}e^{-\lambda_1 |x|} + \fr{A_2}{\lambda_2}e^{-\lambda_2 |x|}.
\end{align}
The coefficients $A_1$ and $A_2$ can be found from the requirement that $D_0(0) = 0$, which follows from the symmetry of the
system [$p_+(0)=p_{-}(0)$],  
in conjunction with the fact that the SSD is normalized, that is,
$\int^\infty_{-\infty}dx~P_0(x) = 1$, yielding 
\begin{subequations}
\label{A1A2}
\begin{align}
\fr{1}{A_1} &= 2\Big( \fr{1}{\lambda^2_1}-\fr{1}{\lambda_1\lambda_2}\fr{b-\lambda_1 T }
{b-\lambda_2 T } \Big),\\
\fr{1}{A_2} &= 2\Big( \fr{1}{\lambda^2_2}-\fr{1}{\lambda_1\lambda_2}\fr{b-\lambda_2 T}
{b-\lambda_1 T } \Big).
\end{align}
\end{subequations}
Notice that $A_1$, $A_2$, $\lambda_1$ and $\lambda_2$ are related by the normaliuzation condition of the total SSD:
\begin{equation}
\frac{A_1}{\lambda_1^2}+\frac{A_2}{\lambda_2^2}=\frac{1}{2}.
\label{normalization}
\end{equation}

\begin{figure}
\centering
\includegraphics[width=1.0\textwidth]{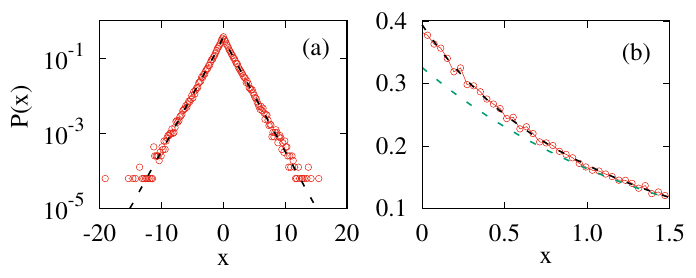}
\centering
\caption{(a) The SSD, $P(x)$, of the RTP obtained from Langevin dynamics simulations (red circles) 
and the analytical result in Eq.~(\ref{px_ss}) (black dashed line). All the model parameters are set to unity: $r = 1$, $v_0 = 1$, $b = 1$, and $T = 1$, 
for which we have (see details in the text): $\lambda_1 \approx 0.69$, $\lambda_2 \approx 2.48$, $A_1/\lambda_1 \approx 0.32$, and $A_2/\lambda_2\approx 0.07$.  
(b) Enlargement of the central region in (a). The green dashed lines depicts
the large $|x|$ asymptotic Laplace distribution $\fr{A_1}{\lambda_1}e^{-\lambda_1|x|}$. Langevin dynamics simulations were conduced with the Euler integration method, with time step $dt=4\times 10^{-3}$ and are based on $10^6$ trajectories.}  
\label{fig1}
\end{figure}

The SSD defined by Eqs.~(\ref{px_ss}) and (\ref{A1A2}), is a normalized linear combination of
two Laplace distributions with distinct length-scales. This form seems reasonable considering that it represents 
the SSD of a particle subjected to two noise sources, thermal and active, which individually lead to steady state single-scale Laplace distributions.
However, the two-noise SSD is not merely a linear combination of the one-noise SSDs: The inverse length scales $\lambda_1$ and $\lambda_2$ are
related to the model parameters characterizing the two noise sources via an intricate relationship expressed by Eq.~(\ref{lmb_k}). 
Fig.~\ref{fig1} shows Langevin Dynamics simulations results corresponding to the case when all the model parameters are set to unity:
$r = 1$, $v_0 = 1$, $b = 1$, and $T = 1$. For this set of model parameters, the
two eigenvalues are $\lambda_1 \approx 0.69$ and $\lambda_2 \approx 2.48$
with the coefficients $A_1/\lambda_1 \approx 0.32$ and $A_2/\lambda_2 \approx 0.07$, i.e., $A_1 \approx 0.22$ and $A_2 \approx 0.17$.
For comparison, the inverse length scales corresponding to Brownian and
RTP without thermal contact dynamics would be, respectively, $\lambda_{\rm Br}=b/T=1$, and $\lambda_{\rm RTP}=b/T^*=2br/(v_0^2-b^2)\rightarrow \infty$ 
(which is a reminder that in the absence of coupling to a heat bath, the RTP moves only when $v_0>b$~\cite{dhar2019run,farago2024confined}). Fig.~\ref{fig1}(a) 
reveals excellent agreement between 
the simulations results (red circles) and the SSD given by Eqs.~(\ref{px_ss}) and (\ref{A1A2}) (dashed-line). Fig.~\ref{fig1}(b) 
shows an enlargement of the central region of the SSD. The green dashed-line shows the Laplace distribution 
$\fr{A_1}{\lambda_1}e^{-\lambda_1|x|}$, which dominates the SSD at large $|x|$. The deviations, on length scales
 not much larger than $\lambda_2^{-1}\sim 0.40$, between the SSD and the asymptotic form  are clearly visible. 

\begin{figure}
\centering
\includegraphics[width=1.0\textwidth]{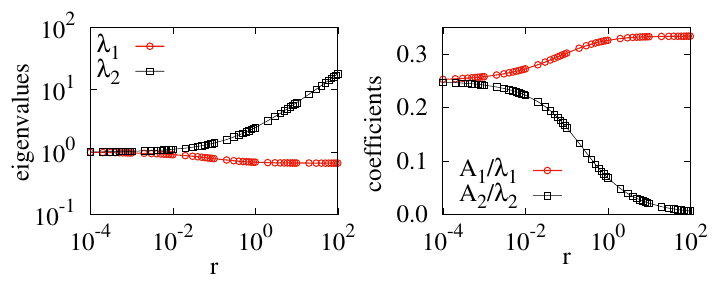}
\caption{(a) The eigenvalues and (b) the normalizations coefficients defining of the SSD,
as a function of the tumbling rate $r$. The model parameter are set to: $b = 1,~T = 1$ and $T_{\rm ac}=v^2_0/2r = 1/2$.}
\label{fig2}
\end{figure}

As mentioned above, in the absence of one of the noises, the SSD takes the form of a single-exponential Laplace distribution. 
Apart from these two particular cases, there is one additional case resulting in a Laplace SSD, which is the thermal limit
of the active noise obtained taking the limits $v_0 \to \infty,~r \to \infty$ such that
the ratio $v^2_0/2r = T_{\rm ac}$. In this limit, the particle is essentially moving under the influence of two independent thermal noises, which is equivalent to Brownian motion with $\lambda = b/(T+T_{\rm ac})$. Fig.~\ref{fig2}(a) shows the dependence of the two eigenvalues
[negative solutions of Eq.~(\ref{lmb_k})] on the rate of the active noise $r$, for $b=1$, $T=1$, and $T_{\rm ac}=1/2$. The smaller $\lambda_1$ varies smoothly from $\lambda_1=b/T=1$ for $r=0$ (no active noise) to $\lambda_1=b/(T+T_{\rm ac})=2/3$, while the larger $\lambda_2$ increases monotonically with $r$, implying that the influence of the second exponential corresponding to the smaller length scale $1/\lambda_2$ becomes limited to the vicinity of the origin. Moreover, Fig.~\ref{fig2}(b), which shows the amplitudes of the two  terms in the SSD, demonstrates that increasing $r$ leads to rapid reduction in the amplitude $A_2/\lambda_2$. For $r\to0$, we have $\lambda_1,\lambda_2\to1$, and the corresponding normalizations are $A_1/\lambda_1=A_2/\lambda_2=1/4$. When $r\rightarrow\infty$ the second exponential becomes irrelevant, and $A_1/\lambda_1\rightarrow \lambda_1/2=1/3$.  

As the properties of the steady state of the RTP are understood, a naturally arising question
is how does the RTP relax to its steady state given the information about its initial
configuration? We proceed to answer this question in the next section.

\section{Time dependent solution}
\label{secPxt}
We now return to the set of Eqs.~(\ref{fpe}) and assume that at $t=0$, the particle is located 
at the origin with equal probability to move to the right or left, i.e., $p_\pm(x,0) = \fr{1}{2}\delta(x)$.  
Transforming Eqs.~(\ref{fpe}) in Laplace space gives
\begin{subequations}
\label{fpe_ls}
\begin{align}
s\tl{p}_+(x,s) &- \fr{1}{2}\delta(x) = -\dd_x [(f(x)+v_0)\tl{p}_+(x,s)]
- r \tl{p}_+(x,s) + r \tl{p}_-(x,s) + T \dd_{xx} \tl{p}_+(x,s),\\
s\tl{p}_-(x,s) &- \fr{1}{2}\delta(x) = -\dd_x [(f(x)-v_0)\tl{p}_-(x,s)]
+ r \tl{p}_+(x,s) - r \tl{p}_-(x,s) + T \dd_{xx} \tl{p}_-(x,s),
\end{align}
\end{subequations}
where $\tl{p}_\pm(x,s) = \int^\infty_0 dt~e^{-st}p_\pm(x,t)$ denotes the Laplace
transforms of $p_\pm(x,t)$. Analogous to the analysis of the steady state in the previous section,
the symmetry of the problem, including the initial conditions, implies that 
Eq.~(\ref{fpe_ls}) need to be solved only for $x \geq 0$, since $\tl{p}_\pm(-x,s)=\tl{p}_\mp(x,s)$.

Also analogous to the derivation of the SSD, we define $\tl{P}_0(x,s)=\tl{P}(x,s)$ to be the Lapalce transform of the SSD, and further introduce the following quantities 
\begin{subequations}
\label{s0d0_ls}
\begin{align}
\tl{P}_0(x,s) &= \tl{p}_+(x,s) + \tl{p}_-(x,s),\\
\tl{P}_1(x,s) &= \dd_x P_0(x,s),\\
\tl{D}_0(x,s) &= \tl{p}_+(x,s) - \tl{p}_-(x,s),\\
\tl{D}_1(x,s) &= \dd_x \tl{D}_0(x,s),
\end{align}
\end{subequations}
that satisfy the set of equations (\ref{fpe_ls}):
\begin{align}
\label{mat4_ls_xg0}
\dd_x\begin{pmatrix}
\tl{P}_0\\
\tl{P}_1\\
\tl{D}_0\\
\tl{D}_1\\
\end{pmatrix}
= \underbrace{\begin{pmatrix}
0 & 1 & 0 & 0\\
\fr{s}{T} & -\fr{b}{T} & 0 & \fr{v_0}{T}\\
0 & 0 & 0 & 1\\
0 & \fr{v_0}{T} & \fr{2r+s}{T} & -\fr{b}{T}
\end{pmatrix}}_{M_4}
\begin{pmatrix}
\tl{P}_0\\
\tl{P}_1\\
\tl{D}_0\\
\tl{D}_1\\
\end{pmatrix}.
\end{align}
In contrast to Eq.~(\ref{mat3_ss}), this matrix differential equation cannot be decomposed into smaller matrices.
For the matrix $M_4$, we have
\begin{subequations}
\label{trdet}
\begin{align}
\text{tr}~M_4 &= -\fr{2b}{T} < 0,\\
\text{det}~M_4 &= \fr{s(s+2r)}{T} > 0,
\end{align}
\end{subequations}
which implies that either all the eigenvalues are negative or two are positive
and two are negative. In the limit $s \to 0$ corresponding to large times, the 
eigenvalues approach their respective values for the SSD, see section~\ref{sec:ssd}. 
One of the four eigenvalues approaches zero in the limit $s \to 0$, and the remaining three 
eigenvalues converge to the eigenvalues of the matrix $M_3$ in Eq.~(\ref{mat3_ss}). 
As one of the eigenvalues of $M_3$ is positive, this rules out the possibility of all four eigenvalues
of the matrix $M_4$ are negative, leaving us with the only option that two eigenvalues are positive
and two are negative. The eigenvalues are obtained from the quartic equation:
\begin{align}
\label{eig4}
\lambda^4 + \fr{2b}{T}\lambda^3 + \Big[-\fr{v^2_0}{T^2} - \fr{2(s+r)}{T}
+ \fr{b^2}{T^2}\Big]\lambda^2 - \fr{2b(s+r)}{T^2}\lambda +
\fr{s(s+2r)}{T^2} = 0.
\end{align}

Denoting by $\lambda_1(s)>0$ and $\lambda_2(s)>0$ the absolute values of the two negative roots of 
Eq.~(\ref{eig4}), the rest of the derivation follows almost  identically to the derivation of the SSD in section \ref{sec:ssd}. 
For $x>0$, we have
\begin{subequations}
\label{s0d0_ls_xg0}
\begin{align}
\tl{P}_0(x,s) &= \fr{B_1}{\lambda_1}e^{-\lambda_1 x} + \fr{B_2}{\lambda_2}e^{-\lambda_2 x},\\
\tl{P}_1(x,s) &=- B_1 e^{-\lambda_1 x} - B_2 e^{-\lambda_2 x},\\
\tl{D}_0(x,s) &= \fr{T}{v_0\lambda_1}\Big(\fr{b}{T}-\lambda_1 + \fr{s}{T\lambda_1}
\Big)B_1 e^{-\lambda_1 x}
+ \fr{T}{v_0\lambda_2}\Big(\fr{b}{T}-\lambda_2
+\fr{s}{T\lambda_2} \Big)B_2 e^{-\lambda_2 x}\\
\tl{D}_1(x,s) &= -\fr{T}{v_0}\Big( \fr{b}{T}-\lambda_1 +\fr{s}{T\lambda_1}
\Big)B_1 e^{-\lambda_1 x}
+ \fr{T}{v_0}\Big( \fr{b}{T}-\lambda_2
+ \fr{s}{T\lambda_2} \Big)B_2 e^{-\lambda_2 x},
\end{align}
\end{subequations}
where the $s$-dependence of the eigenvalues, $\lambda_1$ and $\lambda_2$, and the corresponding coefficients, 
 $B_1$ and $B_2$, has been omitted for brevity. The latter are obtained from the condition that
 $\tl{D}_0(0,s) = 0$, combined with the normalization
of the probability distribution, $\int^\infty_{-\infty}dx~\tl{P}_0(x,s) = \fr{1}{s}$. These yield that
\begin{subequations}
\label{b1b2}
\begin{align}
\fr{1}{B_1} &= 2s\Big(\fr{1}{\lambda^2_1}-\fr{1}{\lambda_1\lambda_2}
\fr{\fr{b}{T}-\lambda_1 + \fr{s}{T\lambda_1} }
{\fr{b}{T}-\lambda_2 + \fr{s}{T\lambda_2}} \Big),\\
\fr{1}{B_2} &= 2s\Big(\fr{1}{\lambda^2_2}-\fr{1}{\lambda_1\lambda_2}
\fr{\fr{b}{T}-\lambda_2 + \fr{s}{T\lambda_2}}
{ \fr{b}{T}-\lambda_1 + \fr{s}{T\lambda_1}}\Big).
\end{align}
\end{subequations}
Taking into account the symmetry of the problem, the time-dependent distribution in Laplace space reads
\begin{align}
\label{pxt_ls}
\tl{P}(x,s) = \fr{B_1(s)}{\lambda_1(s)}e^{-\lambda_1(s)|x|} + \fr{B_2(s)}{\lambda_2(s)}
e^{-\lambda_2(s)|x|}.
\end{align}
In the limit $s \to 0$, this distribution approaches the SSD obtained earlier in Eq.~(\ref{px_ss}). 

In order to complete the derivation, we need to solve Eq.~(\ref{eig4}) for the two negative eigenvalues, and perform an inverse 
Laplace transform to obtain $P(x,t)$ from $\tl{P}(x,s)$ in Eq.~(\ref{pxt_ls}). 
Although Eq.~(\ref{eig4}) can be solved exactly, for example with MAXIMA, a derivation of  
a closed-form simple expression for the time-dependent probability distribution are practically-impossible to obtain 
because of the $s$-dependence of $\lambda_i(s)$ and $B_i(s)$ ($i=1,2$).  
%Thus, we shall resort in the following section to numerical methods for estimating the eigenvalues for specific values of system parameters %\cite{press1992numerical,eaton1997gnu}.
Instead, we will now introduce closed-form approximations for the eigenvalues and the normalization coefficients that capture the correct relaxation behaviors at very small and large times, and verify numerically their accuracy at the intermediate time scales.    

\section{Relaxation to steady state}
\label{secNp}

Since the probability distribution relaxes to the steady state at very large times, it follows that for $s\to 0$, $\lambda_i(s)\to\lambda_i(0)\equiv\lambda_i$, where $\lambda_{i}$ are the corresponding eigenvalues of the $M_3$ matrix (\ref{mat3_ss}). In the opposite limit $s\to\infty$, 
we expect the eigenvalues to behave like $\lambda_i(s)\sim \sqrt{s/T}$, which is the asymptotic form in Laplace space characterizing the relaxation behavior of a Brownian particle: 
%(which are obtained by solving Eq.~(\ref{eig4}) with $v_0=0$ and $r=0$):
$\lambda_{\rm Br}(s)= (b + \sqrt{b^2 + 4sT})/2T\stackrel{s\to \infty}{\sim} \sqrt{s/T}$, and which implies a similar asymptotic behavior in time regime for $t\to0$, i.e., Gaussian spreading
\begin{equation} 
P(x,t)=\frac{e^{-x^2/4Tt}}{\sqrt{4\pi Tt}}+\cdots
%=\frac{\exp\left[-x^2/4Tt\right]}{\sqrt{4\pi Tt}}\left[1+{\cal O}\left(b\sqrt{t/T}\right)\right].
\label{asymptot1}
\end{equation}
The asymptotic similarity with Brownian motion can be understood by noting that at times smaller than the tumbling time, $t\ll r^{-1}$, the particle propagates from the origin subject to action of the thermal and active noises. 
The latter splits equally between the positive and negative directions at $t=0$, and has a diminishing probability to flip its direction when $t\to0$. 
Thus, the total [active - $\sigma(t)$, and thermal - $\eta(t)$] noise can be   
written approximately as the sum of two {\em thermal}\/ white noises, $\eta_\pm(t)$, with constant drift velocities  $\langle \eta_\pm(t)\rangle=\pm v_0$ and variances $\left\langle [\eta_\pm(t^\prime)\mp v_0][ \eta_\pm(t)\mp v_0]\right\rangle=2T\delta(t-t^\prime)$: 
\begin{equation}
\label{noises}
\sigma(t)+\eta(t)\simeq \frac{1}{2}\eta_+(t)+\frac{1}{2}\eta_-(t)\stackrel{t\to 0}{\longrightarrow}\eta(t),
\end{equation}
which manifests the fact that at extremely small times, the active noise
is "washed out" by the stronger delta-correlated thermal noise.

\begin{figure}
\centering
\includegraphics[width=1.0\textwidth]{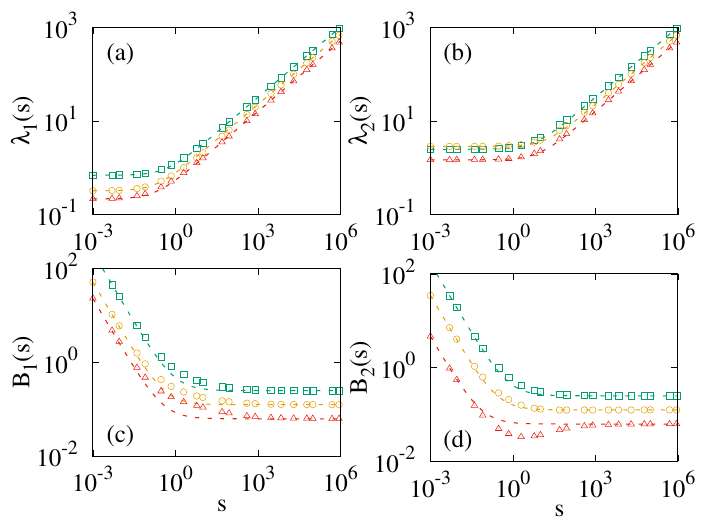}
\caption{(a-b) The dependence of the eigenvalues $\lambda_1$ and $\lambda_2$,
and (c-d) the coefficients $B_1$ and $B_2$, on the Laplace variable $s$. Symbols
indicate the numerically determined values and the dashed lines represent the approximation of Eq.~(\ref{sqrtfit}).
Model parameter values are: $r = 1, T = 1, b = 1, v_0 = 1$ (green, squares);
$r = 4, T = 2, b = 1, v_0 = 3$ (yellow, circles); $r = 3, T = 4, b = 1, v_0 = 2$ (red, triangles).}
\label{fig_lmbd}
\end{figure}

With this in mind, we propose the the following approximation for the eigenvalues in Laplace space:
\begin{align}
\label{sqrtfit}
   \lambda_i(s)\approx\sqrt{s/T+\lambda_i^2},\ ~i = 1,2,
\end{align}
which has the correct asymptotic behavior in the limits $s\to0$ and $s\to\infty$.
Figs.~\ref{fig_lmbd}(a-b) present the dependence of the $\lambda_i(s)$ ($i=1,2$) on the Laplace variable $s$. The values of $\lambda_i(s)$, 
indicated by different symbols of different colors, are obtained by numerically
determining the eigenvalues of the matrix $M_4$ (\ref{mat4_ls_xg0}) for three different sets of models parameters $v_0$, $r$, $T$, and $b=1$. 
As seen in the figures, the agreement between the numerical results and Eq.~(\ref{sqrtfit}) is nearly-perfect
over the entire $s$ range, and not only for $s\to 0$.  The three sets of model parameters displayed in Figs.~\ref{fig_lmbd} correspond to situations in which the active temperature $T_{\rm ac}=v_0^2/2r$ is comparable to the thermal temperature $T$. Full inspection of the quality of this approximation in cases when one of the temperatures is much larger than the other requires the analysis of a 4-dimensional parameter space, which is beyond the scope of the paper.

Along similar lines, we propose the following form for the coefficients $B_i(s)$ in Eq.~(\ref{pxt_ls}) 
\begin{subequations}
\label{num_b1b2}
\begin{align}
&B_i(s) \approx \frac{A_{i}}{s} +\frac{\alpha_i}{T},\ ~i = 1,2\\
&\alpha_1=\alpha_2=\frac{1}{4} \label{num_a1a2}.
\end{align}
\end{subequations}
This form captures the correct asymptotic behaviors in both the small and the large time limits. The first term on the r.h.s.~of Eq.~(\ref{num_b1b2}) ensures that the SSD is obtained for $t\to\infty$ ($s\to 0$), while the second term on the r.h.s.~provides the opposite asymptotic limit, i.e., the coefficient yielding the Gaussian  distribution~(\ref{asymptot1}) at $t\to 0$ ($s\to\infty$). Note that the coefficients $\alpha_i$ must satisfy $\alpha_1+\alpha_2=1/2$ to ensure the normalization of the distribution function at small times. Based on the numerical results in Figures Figs.~\ref{fig_lmbd}(c-d), they are set equal to each other in Eq.~(\ref{num_a1a2}), as the results suggest their equality. It implies that, initially, the probability splits equally between the modes, which is in contrast to the steady state at which the contributions of the modes to the total probability are not equal. In other words, the modes are implicitly coupled since probability is transferred between them.  Figs.~\ref{fig_lmbd}(c-d) reveal that Eq.~(\ref{num_b1b2}) deviates at the intermediate time scales from the numerically-estimated values of 
$B_i(s)$ which were obtained by substituting the numerical values of the eigenvalues $\lambda_{i}(s)$ in Eq.~(\ref{b1b2}). This is not surprising considering that Eq.~(\ref{num_b1b2}) is constructed so as to capture only the asymptotic
short and long time behaviors. In fact, it does not even guarantee that the probability distribution function is normalized to unity except for the two asymptotic limits.  In the following subsection, we use the approximations of $\lambda_i(s)$ and $B_i(s)$ to derive an approximate analytical expression for the time-dependent distribution function $P(x,t)$, which is correct asymptotically at small and large time and provides a decent approximation for the relaxation behavior over the entire time domain.

\subsection{Probability distribution}
\label{pdf}

Using Eqs.~(\ref{sqrtfit}), (\ref{num_b1b2})  in
Eq.~(\ref{pxt_ls}) we write the probability distribution $\tl{P}(x,s)$ in 
Laplace space as
\begin{align}
\label{pxt_ls00}
\tl{P}_{\rm app}(x,s) = \fr{A_1/s+1/4T}{\sqrt{s/T+\lambda_1^2}}
e^{-\sqrt{s/T+\lambda_1^2}|x|}
+ \fr{A_2/s+1/4T}{\sqrt{s/T+\lambda_2^2}}e^{-\sqrt{s/T+\lambda_2^2}|x|},
\end{align}
where the notation "app" serves as a reminder that this form is based on the approximations of $\lambda_i(s)$ and $B_i(s)$.
To invert the above Laplace transform, we split the numerator in
Eq.~(\ref{pxt_ls00}) into the contributions of the $A_i/s$ and $1/4T$ terms.
This leads to the decomposition 
$\tl{P}_{\rm app}(x,s) = \tl{P}_{\rm dec}(x,s)
+ \tl{P}_{\rm rel}(x,s)$,
where the "decaying" distribution, $\tl{P}_{\rm dec}(x,s)$, corresponds the sum of
terms with prefactor $1/4T$, while the "relaxing" distribution, $\tl{P}_{\rm rel}(x,s)$, is associated with the sum
of terms with prefactor $A_i/s$. The 
Laplace inversion of $\tl{P}_{\rm dec}(x,s)$ results in \cite{oberhettinger2012tables}:
\begin{align}
P_{\rm dec}(x,t) =\sum_{i=1}^2 P_{{\rm dec},i}(x,t)=&\fr{1}{2\sqrt{4\pi T t}}e^{-\fr{t}{\tau_1}-\fr{x^2}{4T t}}
+ \fr{1}{2\sqrt{4\pi T t}}e^{-\fr{t}{\tau_2}-\fr{x^2}{4 T t}},
\label{p1xt}
\end{align}
where 
\begin{equation}
    \label{relaxtimes}
    \tau_{i}=(\lambda^2_1T)^{-1},\ ~i = 1,2
\end{equation}
are two time constants characterizing the relaxation dynamics of the two Laplace distributions that constitute the SSD, Eq.~(\ref{px_ss}).
%A look at the structure of $P_{\rm dec}(x,t)$ reveals that at large times it decays exponentially to zero with time constants $\tau_i$ ($i=1,2$). 
For the set model parameters $r = 1, T = 1, b = 1, v_0 = 1$, we find $\tau_i=2.11$ and $\tau_2=0.18$, which means that the mode which dominates the SSD in this case relaxes much slower than the other mode.
For times $t \ll \text{min}\{\tau_1,\tau_2\}$, with
$\text{min}\{,\}$ denoting the minimum of the two arguments, we find
$P_{\rm dec}(x,t) \approx \exp[-x^2/4T t]/\sqrt{4\pi T t} + \cdots$,
consistent with Eq.~(\ref{asymptot1}).
%with the additional terms depending on $\tau_1$ and $\tau_2$. This
%implies that the small time behavior of the distribution $P_{\rm dec}(x,t) \equiv
%{\cal L}^{-1}[\tl{P}(x,s)]$ is a Gaussian (to the leading order).

The term corresponding to $A_i/s$ describes the simultaneous emergence of
and relaxation to the steady state. For the Laplace inversion, we use
${\cal L}^{-1}[F(s)/s]=\int_0^t{\cal L}^{-1}[F(s)](t^{\prime})dt^{\prime}$,
and obtain 
\begin{align}
P_{\rm rel}(x,t) =\sum_{i=1}^2 P_{{\rm rel},i}(x,t)= A_1\int^t_0 dt^{\prime} \sqrt{\fr{T}{\pi t^{\prime}}}
e^{-\fr{t^{\prime}}{\tau_1}-\fr{x^2}{4Tt^\prime}}
+A_2\int^t_0 dt^{\prime} \sqrt{\fr{T}{\pi t^{\prime}}}e^{-\fr{t^{\prime}}{\tau_2}-
\fr{x^2}{4Tt^\prime}}.
\label{p2xt}
\end{align}
For $t\rightarrow\infty$, the two integrals in Eq.~(\ref{p2xt}), 
represent the Laplace transform of a Gaussian (with $1/\tau_i$ playing the role of the variable $s$ for each relaxation mode), which is  
the Laplace distribution~\cite{oberhettinger2012tables}. Thus, 
\begin{align}
\label{hxtinf}
\lim_{t\to\infty}P_{\rm rel}(x,t)= A_1\sqrt{T\tau_1}e^{-|x|/\sqrt{T\tau_1}}+A_2\sqrt{T\tau_2}e^{-|x|/\sqrt{T\tau_2}},
\end{align}
which is the SSD (\ref{px_ss}) because $\sqrt{T\tau_i}=1/\lambda_i$ [see Eq.~(\ref{relaxtimes})].

Recall that the approximate time-dependent distribution function $P_{\rm app}(x,t)=P_{\rm dec}(x,t)+P_{\rm rel}(x,t)$ is not properly normalized. Explicitly, we have that
\begin{equation}
    {\cal N}(t)=\int_{-\infty}^{\infty} dx~P_{\rm app}(x,t) = 1 + \Big(\fr{1}{2}-\fr{2A_1}{\lambda^2_1}\Big)
e^{-t/\tau_1} + \Big(\fr{1}{2}-\fr{2A_2}{\lambda^2_2}\Big)e^{-t/\tau_2},
\label{norma}
\end{equation}
and its easy to verify that ${\cal N}(t)\to 1$ in both limits $t\to0$ and $t\to\infty$. To ensure normalization at all times, we can redefine
\begin{equation}
    P_{\rm app}(x,t)=\frac{P_{\rm dec}(x,t)+P_{\rm rel}(x,t)}{{\cal N}(t)},
    \label{papp}
\end{equation}
with $P_{\rm dec}(x,t)$, $P_{\rm rel}(x,t)$, and ${\cal N}(t)$ given by Eqs.~(\ref{p1xt}), (\ref{p2xt}), and (\ref{norma}), respectively.

\subsection{Mean square displacement}
\label{msd}

\begin{figure}
\centering
\includegraphics[width=1.0\textwidth]{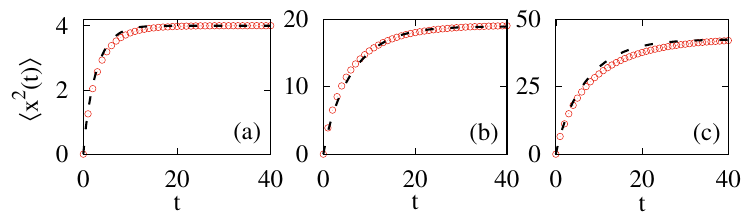}
\caption{MSD, $\lr x^2(t) \rl$, of the RTP 
evaluated from Langevin dynamics simulations 
(red circles) vs.~the predictions of Eq.~(\ref{msd_time}) (black dashed line). 
Model parameter values are (a) $r = 1,\ T = 1,\ b = 1$, and $v_0 = 1$; (b) $r = 4,\ T = 2,\ b = 1$, and $v_0 = 3$;
and (c) $r = 3,\ T = 4,\ b = 1$, and $v_0 = 2$. The black dashed lies shows the 
theoretically estimated MSD [Eq.~(\ref{msd_time})]. Langevin dynamics simulations were conduced with the Euler integration method, with time step $dt=4\times 10^{-3}$ and are based on $5\times10^7$ trajectories. Error bars in the simulation results are smaller than the symbol sizes.}
\label{fig_msd}
\end{figure}

A quantity of practical interest
in context of the dynamics of the RTP is its mean square displacement (MSD), 
\begin{align}
\label{msd_ls}
\lr \tl{x}^2(t) \rl =\int^\infty_{-\infty}dx~x^2 {P}(x,t).
\end{align}
Using $P_{\rm app}(x,t)$ from Eq.~(\ref{papp}) in Eq.~(\ref{msd_ls}), we obtain 
\begin{align}
\label{msd_time}
\lr x^2_{\rm app}(t) \rl = & \left\{\fr{4A_1}{\lambda^2_1}\left[1-\left(1+\frac{t}{\tau_1}\right)
e^{-\frac{t}{\tau_1}}\right] + \frac{t}{\tau_1} e^{-\frac{t}{\tau_1}}\right.\nonumber\\
&\left.+ \fr{4A_2}{\lambda^2_2}\left[1-\left(1+\frac{t}{\tau_2}\right)
e^{-\frac{t}{\tau_2}}\right]
+ \frac{t}{\tau_1} e^{-\frac{t}{\tau_2}}\right\}\nonumber \\
&\times\left[1 + \Big(\fr{1}{2}-\fr{2A_1}{\lambda^2_1}\Big)
e^{-t/\tau_1} + \Big(\fr{1}{2}-\fr{2A_2}{\lambda^2_2}\Big)e^{-t/\tau_2}\right]^{-1}.
\end{align}
Fig.~\ref{fig_msd} shows Langevin dynamics results for the MSD, $\lr x^2(t) \rl$, for the three sets of model  parameters also displayed in Fig.~(\ref{fig_lmbd}). The results are found in 
a very good agreement with Eq.~(\ref{msd_time}), which lends credibility to our approximations. 

\subsection{Local relaxation time}

 At small times and to leading order, Eq.~(\ref{msd_time}) reads $\langle x^2(t)\rangle\approx 2Tt$, suggesting that initially the dominant relaxation mechanism is thermal diffusion. This is anticipated considering the asymptotic behavior of $P(x,t)$ - see Eq.~(\ref{asymptot1}).  Furthermore, from the small time limit  $t\to0$ in Eq.~(\ref{p1xt}) for $P_{\rm dec}(x,t)$, we can conclude that thermal Gaussian spreading governs the relaxation to steady-state on times $t\lesssim\tau_i $ and on length scales $|x|\lesssim \lambda_i^{-1}$. This spatial region constitutes the "core" of the distribution, where the steady state probability is concentrated.   
%This conclusion is consistent with Eq.~(\ref{relaxtimes}), relating the length $\lambda_i^{-1}$  to the square root of the time $\tau_i$ with $T$ being the coefficient of proportionality: $(\lambda_i ^{-1})^2=T\tau_i$. 
The same conclusion can be also inferred from $P_{\rm rel}(x,t)$ in Eq,.~(\ref{p2xt}), from which we can estimate the fraction of the probability density that has already relaxed to the steady state
\begin{equation}
    \Pi_{{\rm rel},i}(t)\equiv\int_{-\infty}^{\infty}  P_{{\rm rel},i}(x,t)dx=2A_iT\tau_i\left(1-e^{-t/\tau_i}\right),
\end{equation}
which approaches saturation exponentially with time scale $\tau_i$. 

At larger times, the relaxation dynamics can be inferred from Eq.~(\ref{p2xt}) for $P_{\rm rel}(x,t)$. The integrals in this equation can be evaluated using the saddle-point approximation. 
%which is applicable at sufficiently large times. 
Explicitly, we define the variable $0\leq u=t'/t\leq 1$ and rewrite 
\begin{equation}
\label{ssa}
P_{\rm rel, i}(x,t) = A_i\sqrt{\fr{Tt}{\pi}}\int^1_0 du~\fr{e^{-t\phi(u)}}{\sqrt{u}},
\end{equation}
where $\phi(u)=[u/\tau_i+x^2/(4Tt^2u)]$.
The maximum contribution to the integral $P_{\rm rel,i}(x,t)$ comes from the minimum of the
function $\phi(u)$ which is at $u_{0,i}=(|x|/t)\sqrt{\tau_i/4T}$ if $x$ and $t$ are such that $u_{0,i}<1$. If $u_{0,i} \geq 1$ then the maximal contribution to
the integral comes from the end of the interval, i.e., $u_{0,i} = 1$. Note that the saddle-point approximation is applicable at large times, when the integrand in Eq.~(\ref{ssa}) is {\em sharply}\/ peaked. At such large times, the $1/\sqrt{u}$ contribution in Eq.~(\ref{ssa}) is negligible. With this in mind we find that the saddle-point approximation yields that:
\begin{enumerate}
 \item  The distribution function relaxes to the SSD for $|x|<2(\sqrt{T/\tau_i})t$:\\
 $P_{{\rm rel},i}(x,t)=(A_i/\lambda_i)e^{-\lambda_i|x|}$.
\item The distribution function takes a Gaussian form for $|x|>(2\sqrt{T/\tau_i})t$:\\
$P_{{\rm rel},i}(x,t)\sim (A_i/\lambda_i)e^{-t/\tau_i}e^{-x^2/4Tt}$.
\end{enumerate}
From (i) we identify a "relaxation front" propagating along the $x$ axis at velocity $v^*_i=2\sqrt{T/\tau_i}$ (which can also be written $v^*_i=2T\lambda_i$ - see Eq.~\ref{relaxtimes}) . The distribution function relaxes to steady state at points that have been passed by the front ($|x|<v^*_it$), and decays as a Gaussian at points not yet passed by the front ($|x|>v^*_it$). Keep in mind that the saddle point approximation is applicable only at times  $t\gtrsim \tau_i$ and distances $|x|\gtrsim \lambda_i^{-1}$. In other words, the relaxation front with constant velocity applies only to the "rare" part of the distribution, i.e., the exponential tail of the SSD. 

\begin{figure}[t]
    \centering
    \includegraphics[width=0.9\textwidth]{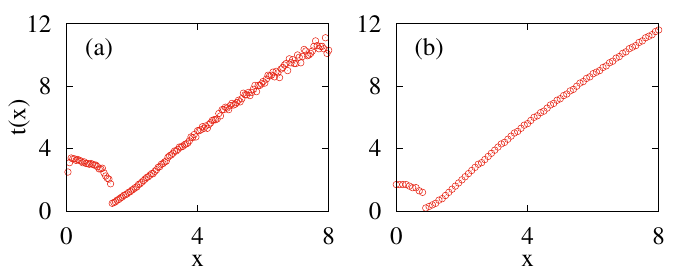}
    \caption{(a) Local relaxation time, $t(x)$, for the RTP obtained by Langevin dynamics simulations with parameters
    $b = 1,~r = 1,~v_0 = 1$ and $T = 1$. (b) The same for a Brownian 
    particle with $b = 1$ and $T = 1$. The RTP results are obtained from Langevin dynamics simulations, while the Brownian motion results are derived from the  exact solution for $P(x,t)$ (see ref.~\cite{chase2016analysis}).  Langevin dynamics simulations were conduced with the Euler integration method, with time step $dt=4\times 10^{-3}$ and are based on $2\times10^6$ trajectories.}
    \label{rlx_times}
\end{figure}

\begin{figure}
    \centering
    \includegraphics[width=0.9\linewidth]{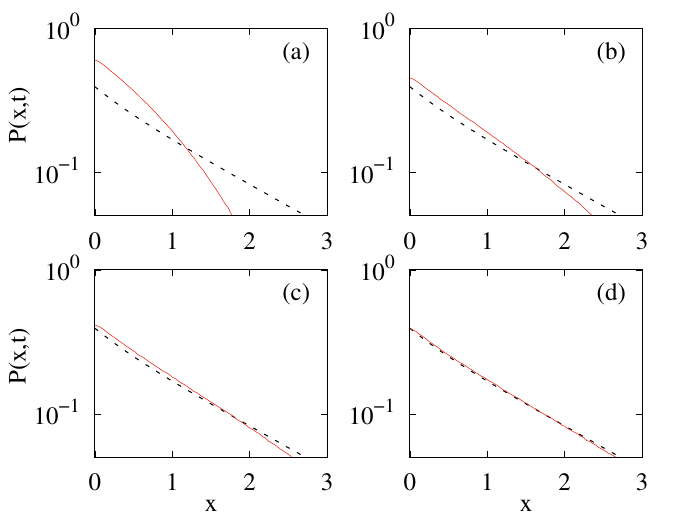}
    \caption{Probability distribution,$P(x,t)$, of the RTP at times (a) $t = 0.5$, (b) $t = 1.5$, (c) $t = 2.5$,
    and (d) $t = 4.5$. Red lines - Langevin dynamics simulations; Black dashed line - the SSD.  Langevin dynamics simulations were conduced with the Euler integration method, with time step $dt=4\times 10^{-3}$ and are based on $10^6$ trajectories.}
    \label{pxt00}
\end{figure}

To summarize the discussion on the nature of the relaxation dynamics at small and large times: Defining the coordinate-dependent {\em local}\/ relaxation time, $t(x)$, as the time at which the distribution function saturates to its steady state value at distance $|x|$ from the origin, 
it is suggested that $t(x)$ levels off to a constant for small $|x|$, and increases linearly for large $|x|$. To test these predictions, we propose an operational definition for $t(x)$, as the time at which the relative difference between the distribution function at $x$ and its steady state value falls below some threshold value $\varepsilon$: $|P(x,t(x))-P(x)|/P(x)<\varepsilon$. Note that $t(x)$, as operationally defined for computer simulations, is an effective time which is not required to coincide with the value of $\tau_i$ in the small $|x|$ regime. Also, $t(x)$ accounts for the combined contribution of the two relaxation modes. However, for the set of parameters used in our simulations ($b = 1,~r = 1,~D = 1$ and $T = 1$), one of the modes is significantly more dominant than the other; thus, $t(x)$ should follow the trends emerging from the single mode analysis, i.e., approach a constant value close to the origin and increase linearly away from it.
Fig.~\ref{rlx_times}(a) shows Langevin dynamics results for $t(x)$ with the relaxation threshold set to $\varepsilon=0.1$. The small- and large-$|x|$ behaviors are indeed observed in the simulation results. Surprisingly, $t(x)$ does not grow monotonically but features a sudden dip at intermediate values of $x$, indicating that local relaxation 
%(to be distinguished from full relaxation of the distribution function to steady state) 
is quickly established at this range. The origin of this distinctive non-monotonic behavior may be understood from the data for $P(x,t)$ plotted against the SSD, $P(x)$, in Fig.~\ref{pxt00} at various times. At small $|x|$, $P(x,t)$ exceeds $P(x)$, which can be ascribed to the outward thermal diffusion of probability from the origin. At large-$|x|$, $P(x,t)$ approaches $P(x)$ from below, and the local relaxation is characterized by the relaxation front. The drop in $t(x)$ at the intermediate range can be attributed to the cancellation of these opposing trends. 
In Fig.~\ref{rlx_times}(b), we plot the local relaxation time of a Brownian particle that starts at the origin and moves in the same confining potential $U(x)=b|x|$ with $b=1$ at temperature $T=1$. The results for $t(x)$ in Fig.~\ref{rlx_times}(b) are derived directly from the exact solution of $P(x,t)$, which is known~\cite{chase2016analysis}. The similarity between the two sub-figures is striking, especially in the nearly-discontinuous drop at the intermediate time scales.

\section{Entropy production at steady state}
\label{entropy}

The rate of entropy production can be calculated exactly for the model system considered herein. For clarity and to explicitly show the dimensionality of the quantities involved, we reintroduced in what follows the Boltzmann constant, $k_B$, and the mobility, $\mu$, which were previously set to unity. At steady state, the rate of entropy production of the RTP in a confiding potential is given by (see section 3.2 in~\cite{entropy})
\begin{equation}
    \dot{S}=\frac{v_0}{\mu k_BT}\int_{-\infty}^{\infty} \left[v_0P(x)-\mu b\,{\rm sgn}(x)D_0(x) \right]dx,
\end{equation}
where $P(x)$ is the SSD Eq.~(\ref{px_ss}) and $D_0(x)=p_+(x)-p_-(x)$ is given in Eq.~(\ref{s1d0d1}). Performing the integrals yields 
\begin{equation}
\dot{S}=\frac{v_0^2}{\mu k_BT}-\frac{2\mu b^2}{k_BT}\left\{\left[1-\frac{\lambda_1 k_BT}{b}\right]\frac{A_1}{\lambda_1^2}
+\left[1-\frac{\lambda_2 k_BT}{b}\right]\frac{A_2}{\lambda_2^2}\right\}.
\label{eprod1}
\end{equation}
 Denoting  the fractional weights of the two modes composing the SSD by $\phi_i=2A_i/\lambda_i^2$ ($\phi_1+\phi_2=1$), Eq.~(\ref{eprod1}) can be written as
\begin{equation}
\dot{S}=\frac{v_0^2-\mu^2b^2}{\mu k_BT}+\mu b^2\left(\frac{\lambda_1}{b}\phi_1
+\frac{\lambda_2}{b}\phi_2\right).
\label{eprod2}
\end{equation}
Eq.~(\ref{eprod2}) expresses the rate of entropy production in terms of the partition of the steady state probability between the modes and their respective correlation-lengths $\lambda_i$. Defining the effective mode temperatures $k_BT_i=b/\lambda_i$ from the equivalent Boltzmann distributions $\exp(-\lambda_i|x|)=\exp(-b|x|/k_BT_i)$,  we have
\begin{equation}
\dot{S}=\frac{v_0^2-\mu^2b^2}{\mu k_BT}+\mu b^2\left(\frac{\phi_1}{k_BT_1}
+\frac{\phi_2}{k_BT_2}\right)=\frac{v_0^2}{\mu k_BT}+\mu b^2\left(\frac{\phi_1}{k_BT_1}
+\frac{\phi_2}{k_BT_2}-\frac{1}{k_BT}\right).
\label{eprod3}
\end{equation}
The second equality in (\ref{eprod3}) suggests that
entropy production at steady state consists of two contributions -  one associated with the energy required for the generation of the active noise source, and the other related to heat exchange between 
the thermal bath and the two mode sub-systems. 
%However, this separation is somewhat misleading or artificial: 
Keep in mind that the simple form of the second contribution may be a bit misleading because the temperatures ($T_i$) and the sizes ($\phi_i$) of the two modes depend on all the model parameters, including $T$ and $v_0$.  
%In the pure active limit, $T\to0$, $\dot{\Phi}\simeq (v_0^2-\mu^2b^2)/\mu k_BT$, which means that the requirement that $v_0\geq\mu b$ emerges from enforcing non-negative entropy production in the system, a direct consequence of the second law of thermodynamics. 

In the limit $v_0\to0$, $r\to0$ with a finite active temperature $k_BT_{\rm ac}=v_0^2/2\mu r$, we have $\phi_1=\phi_2=0.5$, and $T_1=T_2=T$ yielding, $\dot{S}=0$ (see Fig.~\ref{fig2}). Moreover, the SSD in this case converges to the equilibrium Boltzmann distribution $P(x)\sim\exp[-b|x|/k_BT[$, suggesting that the system approaches the pure thermal limit. Interestingly, the influence of such a weak ($v_0\to0$) but persistent ($r\to0$) telegraphic noise does not fade away when the particle is {\em not}\/ subject to a confining potential, i.e. for $b=0$. In free space, the long-time diffusion coefficient is the sum of thermodynamic and active temperatures and is equal to $\mu k_B(T+T_{\rm ac})$~\cite{Malakar_2018}, highlighting the differences between non-equilibrium entropy production in open vs.~confined systems.  In other words, in this limit the configurational temperature defined by the SSD remains the thermal temperature $T$, while the kinetic temperature defined by the diffusion coefficient and via the Einstein relation is $T+T_{\rm ac}$.

\section{Summary and conclusions}
\label{secSS}
We study motion of a run and tumble particle in a piecewise linear confining potential in contact with
a heat bath. The active noise modifies the natural tendency of a Brownian particle to relax to the equilibrium Boltzmann
distribution. We find that the probability distribution $P(x,t)$ of the RTP is described by a mixture of two
distributions, both in transient as well as at the steady state. At short times,
$P(x,t)$ is a mixture of two Gaussian distributions, while at long times it
relaxes to a mixture of two Laplace distributions. The relaxation towards the steady
state $P(x)$ is quite nontrivial. The particle starts at the origin and, thus, the steady state builds from
the center outwards, but the local relaxation time $t(x)$ exhibits a distinctive non-monotonic behavior with a sharp minimum at
intermediate length scales. Close to the origin, at the core of the distribution,
thermal diffusion dominates the relaxation dynamics, while at the tails
it is established at times that grow linearly with $|x|$.
%with $P(x,t)$ showing Gaussian decay outside of the  spatiotemporal region defined by the propagating relaxation front. 
The propagation of the relaxation front in the tail of the distribution is governed by exponentially rare stochastic trajectories that exhibit persistent motion in a single direction, with characteristic durations that scale linearly with $|x|$. This relaxation behavior is also observed in the case of pure Brownian motion, i.e., in the absence of active noise. However, the presence of active noise qualitatively alters the distribution, leading to a splitting into two distinct modes. These modes are characterized by effective temperatures, relaxation times, and front propagation velocities, all of which depend in a non-trivial and intricate way on the model parameters, including the thermodynamic temperature of the surrounding heat bath.

The detailed analysis presented in this work is limited
to the system in a Markovian setting. This, however, is a sort of 
an idealization as
independence of tumbling times may not always hold in practice, for example, for
swimming bacteria \cite{detcheverry2017generalized}. This also brings us within the
realm of heterogeneous diffusion processes with inclusion of memory effects
\cite{sandev2022stochastic,sandev2024fractional}. We shall explore these non-Markovian
effects~\cite{farago2024confined} and related properties in future works. We will also extend our investigations to other forms of 
confining potentials, including these where the steady state of the pure RTP is 
defined only over a finite support, and where constant active temperatures $T_i$ that are associated with Botzmann-like distributions are not straightforwardly defined.  Finally, it would be also interesting to expand our study to dimensions higher than one. While exact solutions of RTP dynamics are typically hard to obtain in two- and three-dimensions, in cases where there is spherical symmetry, the SSD may have similar characteristics to the one-dimensional solution because the orientational moves can be mapped onto a one-dimensional model with different statistics of tumbling lengths~\cite{Elgeti_2015,smith23}. 

\vspace{1cm}
\textit{Acknowledgments}:
We thank Naftali Smith for useful discussions. RKS thanks Kreitman Fellowship for
Financial support and Eli Pollak for discussion about Laplace inversions.

\vspace{1cm}
\noindent{\bf References}
\vspace{0.5cm} 
\bibliographystyle{unsrt}
\bibliography{rtp}

\end{document}